# An Empirical Study of Path Feasibility Queries

Asankhaya Sharma
Department of Computer Science
National University of Singapore
asankhs@comp.nus.edu.sg

*Abstract* — In this paper we present a comparative study of path feasibility queries generated during path exploration based software engineering methods. Symbolic execution based methods are gaining importance in different aspects of software engineering e.g. proving properties about programs, test case generation, comparing different executions of programs. These methods use SMT solvers to check the satisfiability of path feasibility queries written as a formula in the supported theories. We study the performance of solving such path feasibility queries using SMT solvers for real world programs. Our path condition formulas are generated in a theory of quantifier free bit vectors with arrays (QF_ABV). We show that among the different SMT solvers, STP is better than Z3 by an order of magnitude for such kind of queries. As an application we design a new program analysis (Change Value Analysis) based on our study which exploits undefined behaviors in programs. We have implemented our analysis in LLVM and tested it with the benchmark of SIR programs. It reduces the time taken for solving path feasibility queries by 48%. The study can serve as guidance to practitioners using path feasibility queries to create scalable software engineering methods based on symbolic execution.

## I. INTRODUCTION

In recent years, several path exploration based methods have been proposed for analysis and testing of programs. These methods make use of SMT Solvers to check for satisfiability of path condition formulas. We are interested in studying the performance of modern SMT solvers for these kinds of queries. SMT solvers use specialized procedures for different theories along with SAT solving to check for satisfiability of formulas. SMT solvers can also be used to check verification conditions generated during safety property checking of programs. However the formulas generated during verification typically test the expressivity of the SMT solver, as different programming constructs give rise to formulas from various theories like arrays, bit vector, uninterpreted functions etc. These formulas can also contain quantifiers; in the presence of quantifiers for most theories supported in current SMT Solvers (like Z3 [3]) the procedure for checking satisfiability is semi-decidable. Hence a SMT solver can timeout or give an unknown response to such a query.

On the other hand, the formulas generated from path condition calculation test the scalability of the SMT solver. These formulas can be large since they capture the entire path taken by an input (symbolic or concrete) through the program. Also these formulas can be expressed in a quantifier free fragment of a theory (e.g. QF_ABV) supported by the SMT solver. Our study describes the performance of these kinds of path feasibility queries on current SMT solvers.

We take an observational approach and compare existing SMT solvers (STP [2] and Z3) with off the shelf tools (Bitblaze [1] and Klee [4]) to generate path feasibility queries for real world C programs (libPNG [7] and SQLite [5]). The study focusses on comparing different program constructs within a program as well as comparing different kind of programs to ascertain if any of these lead to problems while solving the formula using SMT Solver. It is important to define what we mean by problems for SMT solvers. Solving a large formula is expected to take more time, what we mean by a problem (or anomaly) is if the time taken by the SMT solver is unexpectedly more or less. A similar sized formula generated from a different program may take more or less time when compared to the unexpected formula. We are interested in such formulas as they can help take into account different optimizations while doing path exploration. Which can in turn lead to an increase in scalability of path based program analyses. The main contributions of this study are as follows.

- Our experiments show that for path feasibility queries in the fragment of QF_ABV STP performs better than Z3 in general by an order of magnitude.
- Comparing different formats of formulas we see that SMTLIB2 format [8] is much more concise compared to the STP's format (up to 2 times) and this helps while solving large formulas.
- Various compiler optimizations affect the efficiency of symbolic execution. Based on our study we propose a new program analysis (Chang Value Analysis) which reduces the time taken to solve path feasibility queries for SIR programs by 48%.
- The formulas generated by our study can serve as a benchmark [12] for different symbolic execution engines and constraint solvers to evaluate their performance.

In the next section we describe some background of the tools used for this study and the experimental setup. In section III we detail our methodology for carrying out the experiments, section IV covers the experiments themselves and the data we collected. In section V and VI we discuss our results and an application; in section VII we mention some threats to the validity of this study. Section VIII covers some related work and finally we conclude in section IX with some pointers for future work.

## II. BACKGROUND AND SETUP

At the beginning of this study, we were faced with a number of choices for tools we can use in both path condition generation and subsequent satisfiability checking. The tools we considered for generating path feasibility queries were Klee, Bitblaze and Java Path Finder (JPF). All the three tools have different use cases and are already used by existing test generation and analysis methods. Based on our use of the tools we found that Klee works well for C programs which are self-contained. Klee compiles the source code of a C program into LLVM bit code which forms the basis of further analysis. This assumes we have access to all the source code. Any external library or system call has to be modeled separately. This limits the usefulness of Klee, as we cannot directly analyze binaries using it. Klee [4] comes with support for POSIX and other basic Linux system commands already modeled in to mitigate the external function call problem.

The Bitblaze binary analysis system has two different components, the Temu component [1] can be used to analyze binaries and collect execution traces, while the Vine component [1] can generate path feasibility queries using these traces. The formula generated by Bitblaze contains constraints on the assembly code which have to be mapped back to the variables in the original source code of the program. The benefit of using Bitblaze is that we can analyze even programs for which we do not have access to the complete source code and we do not have to worry about external system calls. The external system calls and memory accesses are already handled by Bitblaze. JPF is similar to Bitblaze in the sense that it works on Java byte code but like Klee the external system and library calls have to be modeled in explicitly.

Based on all these considerations we decided to use Klee and Bitblaze for our study initially, but we ended up with using only Bitblaze in the end as Klee doesn't give a way to generate the full path feasibility query. The formulas returned by Klee contain only the constraints on the symbolic input. This is an artifact of way environment and external calls are modeled by Klee. Klee also supports concolic execution in which some parts of the formula are concretized. Hence the constraints on only the symbolic input cannot be compared directly with a path feasibility query from Bitblaze. Hence all the experiments described in this paper are based on path conditions generated by Bitblaze (except for section VI where we use Fuzzgrind [10]). The formula generated by Bitblaze uses a theory of arrays and bit vectors. This limits the choice of different SMT solvers as we need to have a SMT solver which supports this particular fragment (QF_ABV).

Bitblaze generates the formulas in STP format on the other hand Klee and JPF use the CVC format. To compare different SMT solvers we decided to include Z3 in our study along with STP. Z3 is an industrial strength SMT solver in development from Microsoft Research [3]. Z3 supports several theories while STP solver has support for only bit vectors and arrays. Z3 has already been used in many verification tools for checking satisfiability of verification conditions. Hence for this study we choose to focus on STP and Z3 solvers. Since SMTLIB2 is a standard format accepted by both STP and Z3, this also allows us to compare SMTLIB2 with the format generated by Bitblaze.

The programs used for this study are taken from real world C software. We chose SQLite, a popular self-contained database and libPNG, an image manipulation library. As a control for comparative studies between different programs we also include a generic data structure library written in C (GDSL). All the tools and software used in the study are available freely from the web. Our experimental setup uses a standard desktop machine with Intel Core2 Quad @ 2.83 GHz processor and 4 GB of Ram. The operating system used for running the experiments is Ubuntu 10.04. We got the source code of all the software from the web and built it on our machine using GNU C Compiler and the standard OCAML compiler. Z3 is a commercial product for which the source code is not available. We used the Z3 Linux binaries provided by Microsoft for non-commercial use. The experiments are carried out one at a time with no other system resource intensive process running in the background. In the next section we describe our general methodology for conducting the experiments for the study using this setup.

## III. METHODOLOGY

In our experiments we follow an empirically approach, wherein we observe the following general process.

- Collect the execution trace for some sample input given to the program
- Generate the corresponding path condition formula using the trace
- Solve the formula by representing it as a path feasibility query to a SMT Solver

During the process we collect various timing information of each phase using time system command. As an example consider the GDSL library, our sample input is insertion into a map data structure. We launch the program in a virtual machine using Temu (Bitblaze) and load the trace collection plugin. Then we mark the file containing the input as tainted using the *taint_file* command (Bitblaze). This tells Bitblaze to treat that input as symbolic and track it though the program. Running the program then produces a trace file. This trace file is processed by *app_replay* utility (Bitblaze) to generate the path condition corresponding to that input.

This path condition can then be sent to a SMT solver as a path feasibility query for checking satisfiability. Since we are interested in the performance of the solver we measure the time taken for the query by the solver. We modified the *app_replay* utility to generate the variable mappings between the variables used in the formula to the variables used in the assembly code of the program. Using GNU Binutils we can map the symbols used in assembly code to the actual C source code. All our experiments follow this general method of collecting traces, generating formulas, solving queries, measuring time and tracing back to source code if any anomalies are observed. However, in order to compare different program constructs within a program with respect to the contribution they make to the path feasibility query, we modify this method to the following.

- Collect the execution trace for a given input to the program
- Generate the corresponding path condition formula using the trace
- Using the information about control flow points within the program; divide the formula into a series of formulas of increasing size so that the difference between any two consecutive formulas is of almost equal size
- For each of the formulas above solve the path feasibility query using SMT Solver

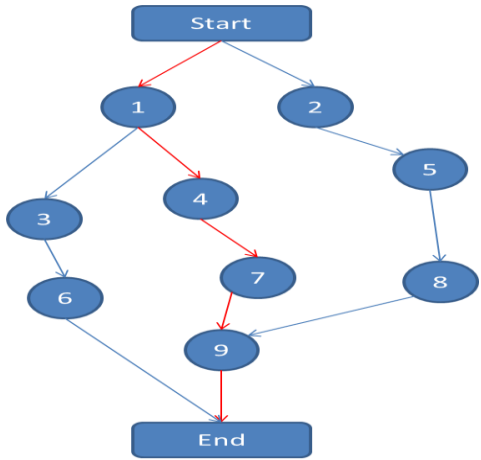

Fig. 1. Control Flow of a Program

In order to better illustrate this method consider Fig 1 representing the control flow of some arbitrary program. The start and end control locations are marked with a rectangular node while the other intermediate points are marked as circular nodes. A given input will trace a single path through this program which is marked in red (Start, 1, 4, 7, 9, End). The path condition formula generated using the trace represents this single path. In order to compare different constructs in the same program, we split this formula into a series of formulas shown in Fig 2. Each node represents some basic block and we try to increase the size of the series of formulas evenly by adding equal number of blocks at a time. We can do this because the trace contains the formula on the assembly language code which has a lot more control locations (round nodes) for us to divide them evenly among the various formulas.

Note that we cannot just split the given formula into equal sections; the formula represents a path condition from start to the end node. And we must take this into account while dividing it into different sections. We use this method for intra-program performance analysis. We measure the time taken for solving each of this series of formulas and by calculating the difference between the consecutive queries we can check for any anomalies.

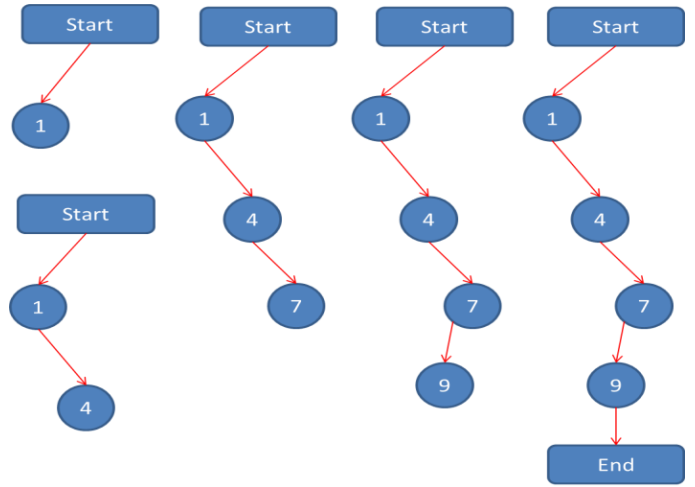

Fig. 2. Different Sections of the Given Path

The technique described above can work for comparing different constructs used within a program, for comparing between different programs we use the following method.

- Collect the execution traces for some sample input to the programs
- Divide the traces from the programs into a series of traces of increasing size so that the difference between any two consecutive traces is equal in size
- Generate the corresponding path conditions for traces of the same size from different programs
- For each of the formulas above solve the path feasibility query using SMT Solver

The above method can be used for inter-program performance analysis by measuring the time taken by the SMT solver to check for satisfiability of a path feasibility query generated from a trace of equal size but from a different program. This method is similar to the intra-program method but instead of the path condition formula we divide the execution trace into different sections as paths in different programs are not going to have similar constructs anyways. The methodology described in this section allows us to carry out various experiments to compare the performance of path feasibility queries. The next section describes the experiments we conducted.

IV. EXPERIMENTS

The experiments we describe are divided into four sections; initially we focus on comparing different constructs in a single path within the program, then we compare different paths in the program and finally we compare different programs along with various compilers. The particular approach used to generate the respective formulas for these experiments is based on the methodology described in section III. At first we describe the analysis based on different sections within the same program for a particular path.

## A. Intra-Program Intra-Path

We conducted two experiments comparing intra-program intra-path performance – for GDSL and SQLite. We could not do the same for libPNG as Bitblaze ran out of memory on our system for even the smallest execution traces we generated for a program using libPNG. Table I shows the data we collected for GDSL. We record the time taken (in seconds) for solving the path feasibility query using both STP and Z3 solvers.

TABLE I. INTRA-PROGRAM INTRA-PATH PERFORMANCE ANALYSIS FOR GDSL

| Name | Time (STP) | Time (Z3) | Time Diff (STP) | Time Diff (Z3) |
|---|---|---|---|---|
| lllist_tr1_1.stp | 0.004 | 0.021 | | |
| lllist_tr1_2.stp | 0.006 | 0.02 | 0.002 | -0.001 |
| lllist_tr1_3.stp | 0.007 | 0.021 | 0.001 | 0.001 |
| lllist_tr1_4.stp | 0.007 | 0.018 | 0 | -0.003 |
| lllist_tr1_5.stp | 0.01 | 0.021 | 0.003 | 0.003 |
| lllist_tr1_6.stp | 0.012 | 0.019 | 0.002 | -0.002 |
| lllist_tr1_7.stp | 0.014 | 0.022 | 0.002 | 0.003 |
| lllist_tr1_8.stp | 0.016 | 0.023 | 0.002 | 0.001 |
| lllist_tr1_9.stp | 0.018 | 0.023 | 0.002 | 0 |
| lllist_tr1_10.stp | 0.022 | 0.022 | 0.004 | -0.001 |
| lllist_tr1_11.stp | 0.02 | 0.025 | -0.002 | 0.003 |
| lllist_tr1_12.stp | 0.024 | 0.026 | 0.004 | 0.001 |
| lllist_tr1.stp | 0.022 | 0.023 | -0.002 | -0.003 |
| lllist_tr1_1.stp | 0.004 | 0.021 | | |

Notice the last column "Time Diff" which records the difference between solving consecutive queries of almost equal size, we find that the values in these columns are almost same for both STP and Z3. This shows that there is very little variability in the time taken to solve the path feasibility query for different sections of the path in the program. Same can be depicted better visually using the graph in Fig 3.

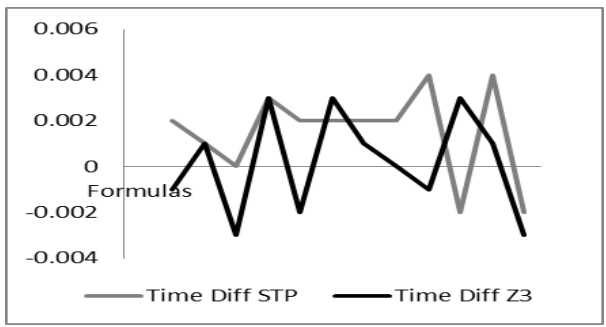

Fig. 3. Graph between Time Difference and Size of formulas for GDSL

As the trace used for these experiments is very small, the timings in these graphs are very small. In order to avoid that we generated several such traces for SQLite of different sizes by varying the input. The complete data collected for all the traces is available online at [12]; here we show only the following four graphs (in Fig 4) based on the data.

The vertical axis in the graphs given in Fig 4 represents the time difference between consecutive formulas. As is clear from the graphs the grey line remains flat (or shows no significant variability). This demonstrates that within a program there is not much difference in terms of solving path feasibility queries. The black line which represents Z3 actually shows a clear rise in the last 2 graphs which we discuss in the section V with results in more details.

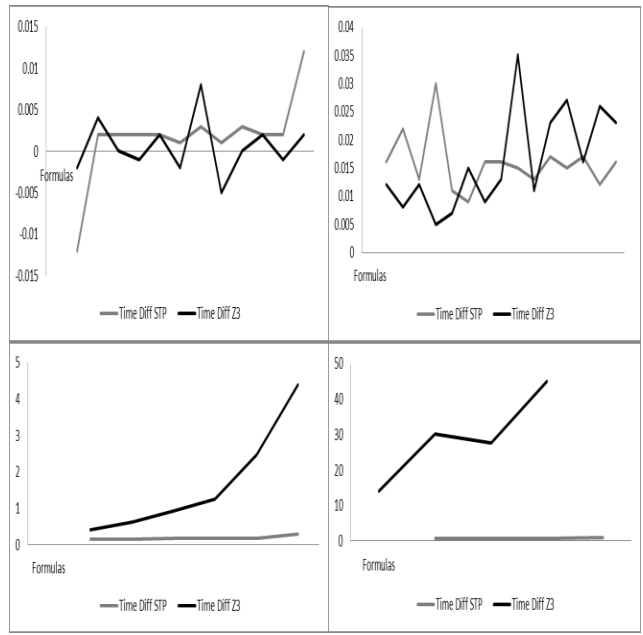

Fig. 4. Graph between Time Difference and Size of formulas for SQLite

Our experiments are performed at different levels of granularity for the path feasibility queries. The vertical axis is almost a magnitude higher in each of the graphs. This captures the performance at different levels, since if we divide the sections in the path condition formula to be too small they are bound to show some variability (which is shown by graph 1 and 2 in Fig 4 for SQLite and in Fig 3 for GDSL). But based on our analysis we find that as we go higher the difference between the consecutive sections of the formulas doesn't change much. Since our intra-program intra-path analysis is unable to find any difference between the performances of path feasibility queries we conduct experiments to compare different paths.

## B. Intra-Program Inter-Path

Based on the performance of Z3 and STP in experiments for SQLite in the previous section we notice that Z3 seems to perform worse for larger queries. In order to check if this is indeed the case we compare the performance of STP with Z3 for different path feasibility queries. We generated traces of increasing size which give rise of corresponding path condition formulas of increasing size. We then use Z3 and STP to solve these queries. We show the data we collected for this experiment in Table II.

TABLE II. INTRA-PROGRAM INTER-PATH PERFORMANCE OF SMT SOLVERS WITH SQLITE

| Trace Name | Time (STP) | Time (Z3) |
|---|---|---|
| sqlite_tr1 | 0.042 | 0.027 |
| sqlite_tr5_test | 0.284 | 0.27 |
| sqlite_tr5 | 1.336 | 10.242 |
| sqlite_tr4 | 3.759 | 119.209 |
| sqlite_tr7 | 45.5 | 1130.494 |
| sqlite_tr6_test | 54.065 | 1249.352 |

The trace name column lists the name of all the traces we used, the traces with names ending in "test" were truncated from the original trace generated by Bitblaze as they contained some special instructions which were not supported by the component in Bitblaze which generates the path feasibility queries from traces. All these traces are collected from SQLite program and are in increasing order of sizes. As seen in Table II Z3 performs worse than STP. Due to this performance gap between STP and Z3, for all the subsequent experiments we use only the STP solver. We discuss this in more detail in section V with results.

To compare the performance of path feasibility queries for different paths we look inside the path condition formula. The formula generated by Bitblaze uses a single array named *mem_arr* to capture the memory reads and writes done by the program. The rest of the free and temporary variables used in the formula are only from the theory of bit vectors. Hence to characterize the effect of control and data flow between the two paths within the same program we count the occurrences of IF-ENDIF constructs and Array Writes in the formula. SMT solvers can directly support if-then-else like statements using the IF-ENDIF construct which is used by Bitblaze. Since the formula is generated from the binary this represents the control flow within the assembly code of the program. Array Write used by Bitblaze in the formula can be used to mark data flows as any tainted input is first copied into memory and then used by the program. Using this information we calculate the counts of these constructs in the formulas generated for different inputs to the SQLite program.

TABLE III. INTRA-PROGRAM INTER-PATH PERFORMANCE ANALYSIS FOR SQLITE

| Name | Lines | IF ENDIFs | Array Writes | IF ENDIF / Array Writes |
|---|---|---|---|---|
| sqlite_tr5 | 758355 | 7052 | 10889 | 0.647 |
| sqlite_tr4 | 2051540 | 23997 | 27519 | 0.872 |
| sqlite_tr7 | 9294739 | 104030 | 125684 | 0.827 |
| sqlite_tr6_test | 10741258 | 118503 | 138054 | 0.858 |

The data collected for this experiment is shown in Table III. We also list the number of lines in the formula as a rough metric for size of the formula. As we can see from Table III, the ratio of number of occurrences of IF-ENDIF to Array Writes doesn't change much for a given program for different paths. We use this ratio to represent the fraction of the path feasibility query that comes due to the control flow in the program as compared to the data flow. The last column lists this ratio for traces captured from SQLite; as shown in Fig 5 this value doesn't change much with increase in the size of formula.

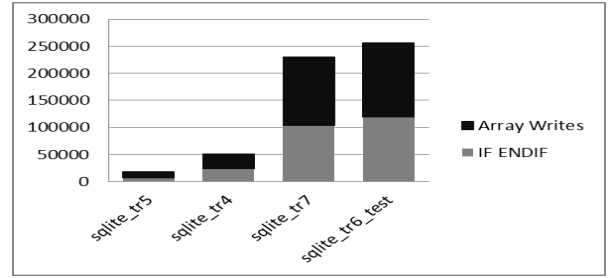

Fig. 5. Graph showing IF ENDIF/Array Write for increasing size of formulas

We use this ratio to characterize the contribution of control and data flow of the program towards the path feasibility query. Since we find that this ratio doesn't change by a significant amount for different paths in the SQLite program it is also used in the inter-program performance analysis. In the next sub-section we describe the inter-program experiments which we carried out. The usefulness of this ratio will be clear from these experiments, we discuss about the implications of using this ratio and its relation with control and data flow of the program in detail in section V with results.

*C. Inter-Program*

In order to compare the performance of different programs we use the ratio defined in the previous sub-section. We generate the required traces of increasing size for libPNG and SQLite as described in section III. The Table IV details the data collected for this experiment. We use trace size and formula size as a metric to judge the difficultly of the path feasibility query generated for that program. The SMT solver used for these experiments is STP. Both libPNG and SQLite show increase in the time taken to solve the queries with increase in trace and formula size. Also note that the IF ENDIF/Array Write ratio is different for the two programs but within each program it doesn't change much with increasing trace or formula size. The ratio is twice as much for libPNG when compared to SQLite. This is discussed further in section V with results.

TABLE IV. INTER-PROGRAM PERFORMANCE ANALYSIS FOR LIBPNG AND SQLITE

| libPNG | | | SQLite | | |
|---|---|---|---|---|---|
| Size | Time | IF ENDIF/Array Writes | Size | Time | IF ENDIF/Array Writes |
| 16.9 | 1.75 | 1.669 | 21.3 | 2.05 | 1.055 |
| 35.8 | 4.93 | 1.778 | 41.3 | 4.04 | 0.884 |
| 57.5 | 14.17 | 1.759 | 59.5 | 6.02 | 0.839 |
| 79.8 | 35.05 | 1.75 | 77.2 | 9.76 | 0.884 |
| 102 | 66.41 | 1.744 | 94 | 16.3 | 0.88 |

While conducting the various experiments we generated formulas for Z3 by converting it from STP format, as Bitblaze outputs the path conditions only in that format. What we found was that the SMTLIB2 format [8] used by Z3 was much more concise. In order to compare the difference between the two formats we used the STP to solve the same formula given in both the formats and collected the time taken. Table V shows the data for these experiments (Mem stands for Memory consumed in MB and Time is in Seconds). Along with the path feasibility formulas generated by Bitblaze for SQLite, we also included four large formulas from original benchmarks for STP [2]. This is the only experiment where we use formulas from existing benchmarks.

TABLE V. INTER-PROGRAM PERFORMANCE ANALYSIS OF STP WITH DIFFERENT FORMATS

| Formula Name | STP Format | | | SMTLIB2 Format | | |
|---|---|---|---|---|---|---|
| | Lines | Time | Mem | Lines | Time | Mem |
| sqlite_tr5 | 758355 | 1.29 | 47 | 6144 | 0.36 | 18 |
| sqlite_tr4 | 2051540 | 3.66 | 82 | 16082 | 0.99 | 19 |
| sqlite_tr7 | 9294739 | 45.1 | 305 | 65563 | 11.2 | 35 |
| sqlite_tr6_test | 10741258 | 53.83 | 585 | 70583 | 13.51 | 40 |
| testcase20 | 2962272 | 37.55 | 680 | 375469 | 35.33 | 751 |
| thumb-noarg | 6445260 | 52.53 | 768 | 912747 | 48.52 | 869 |
| thumb-spin1 | 9801299 | 61.67 | 907 | 1223527 | 53.92 | 1115 |
| thumb-spin-1-2 | 10411398 | 96.37 | 1179 | 1446674 | 80.61 | 1328 |

For the first four formulas which are from SQLite program we see that there is a considerable improvement when using the corresponding formula in SMTLIB2 format. In case of the last four really large formulas taken from [2] we see that there is still some improvement in timings but at the expense of consuming more memory. This shows that a tradeoff has to be made between memory use and time when using a different format. This also points to potential redundancy in the way path feasibility formulas are generated by Bitblaze.

*D. Across-Compiler*

In order to evaluate the effect of using different compliers and optimizations we also conducted an experiment of generating path feasibility queries for different compilers (GCC, Open64 and LLVM). The Bitblaze system was unable to work with binaries produced by Open64 and LLVM compiler (with optimizations turned on). Hence, for these experiments we used another symbolic execution engine Fuzzgrind [10]. To test different compiler optimizations across various kinds of programs we used the benchmark from Software-artifact Infrastructure Repository (SIR) [12]. Fuzzgrind internally uses STP to solve the generated path feasibility queries so we used the same solver to test across compilers. Compilers like GCC offer several coarse optimization levels (-O1, -O2, -Os, etc.) We found that optimizing for space (-Os) helps in producing a binary which is smallest in size and leads to a 5-10% reduction in the time taken to solve the generated path feasibility queries from the application (SIR programs).

Moreover as we describe in section VI by using a simple program analysis before applying these optimizations we can on average reduce the time taken to solve the generated path feasibility queries by 48%. We are not aware of any previous study which describes the effect of compiler optimizations on path feasibility queries. In the next section we evaluate the results of these experiments empirically.

## V. RESULTS

In this section we discuss the results of the study based on the experiments conducted. We also highlight three key findings. The findings are based on the data collected from the experiments as described in the previous section. As our first result, we present a comparison between the two SMT solvers we used – STP and Z3.

*A. Comparing STP and Z3 SMT Solvers*

The Fig 6 depicts the time taken by Z3 and STP for solving path feasibility queries. It is clear that STP is faster in almost all the cases. Note that the vertical axis is actually on log scale and the performance gap between STP and Z3 is a lot. This result also corroborates the results on the performance of Z3 and STP in the SMT competition [9]. Even though Z3 is more expressive and can support many more theories than STP, its performance in the QF_ABV fragment is not as good. Also the procedure implemented in STP [2] is optimized for path feasibility queries so some performance gap between Z3 and STP is expected.

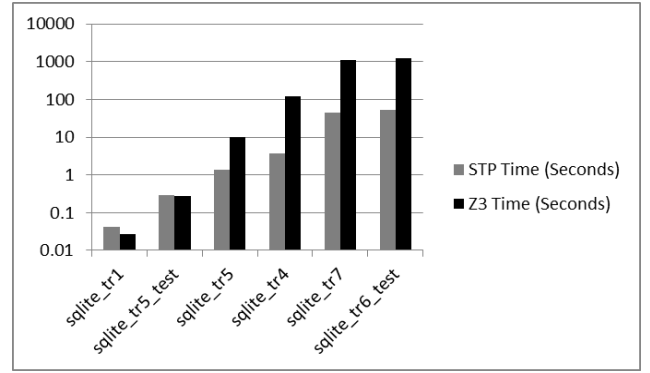

Fig. 6. Comparing STP and Z3

Our results indicate a much bigger gap between performance of Z3 and STP (about 10 times); it can be further explained by the fact that Z3 supports existential theory of arrays while in STP the theory of arrays is non-existential. Even though none of the formulas generated by Bitblaze uses the fact that the array is existential Z3 treats it as such and thus appears to be slower in solving the path feasibility query. Because of the more expressive nature of Z3 it has been known to perform well in checking satisfiability of verification conditions, however based on our results it is not a good choice for solving path feasibility queries (since it is too expressive). This would indicate that for use with path exploration based software engineering methods STP solver would be a better tool.

*B. Comparing STP and SMTLIB2 format using STP Solver*

Our next finding is related to the format used for generating the path feasibility queries. What we have found is that the SMTLIB2 format for SMT solvers is much more concise. It can be seen from the following graph in Fig 7. We compare the time taken by STP to solve the same formula when presented in different formats. We observe that the formula given in

SMTLIB2 format takes much less time in all the cases. We also looked into the statistics generated by STP for each of the formulas and found that a significance amount of time is spent in parsing the file for larger formulas. Since the same formula when written in SMTLIB2 format take lesser space the overall time for the solving the query is reduced.

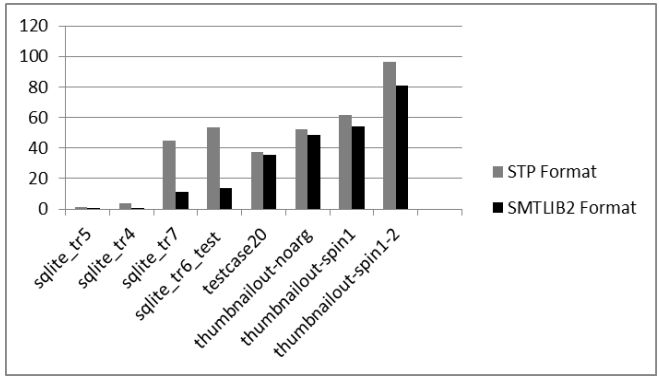

Fig. 7. Comparing STP and SMTLIB2 Format

In this case we also compare some other benchmarks (last four in Fig 7) based on STP format given in [2]. We found that even for those benchmarks which are not generated from path feasibility queries, the SMTLIB2 format gave better performance (although the difference was not as much). This finding indicates that if Bitblaze can generate the path condition in SMTLIB2 format it would be much faster to solve using SMT solvers. Also based on comparison with other existing benchmarks we see that there appears to be some redundancy in formulas corresponding to path feasibility queries. A possible optimization for generating path feasibility query can be to look for ways to reduce this redundancy. If we can generate a smaller and more concise formula by avoiding the introduction of unnecessary intermediate propositions it can lead to better performance.

### C. Comparing libPNG and SQLite using STP Solver

Finally we present a comparison of path condition queries generated from different programs (SQLite and libPNG). The graph in Fig 8 shows the time taken for solving a formula which is generated by the trace of a given size. We see that as the trace size is increased from 1 to 5 MB, the time taken to solve the corresponding path feasibility query also goes up. All the times shown here are for the STP solver since we found Z3 to be too slow for such comparison.

We notice that the increase in time is linear for SQLite but appears exponential for libPNG. It seems to indicate that the STP solver doesn't scale well for path feasibility queries of libPNG traces. In order to explain this result we take help of the ratio between IF ENDIFs and Array Writes which we defined in section IV C. This ratio for libPNG (~1.6) is almost twice as that of SQLite (~0.8), which correlates with worse performance for solving the queries. In the libPNG program there is twice as much contribution from control flow towards the path feasibility query as compared to the SQLite program. This causes difficulty for the STP solver while checking satisfiability of such a formula. This indicates that considering just the data flow dependencies or control flow dependencies within the program is not enough, we need to consider both the data flow and control flow dependencies for a path to characterize the difficulty of the generated path condition formula. This ratio attempts to capture this intuition by quantifying the contribution of data flow and control flow to the given path.

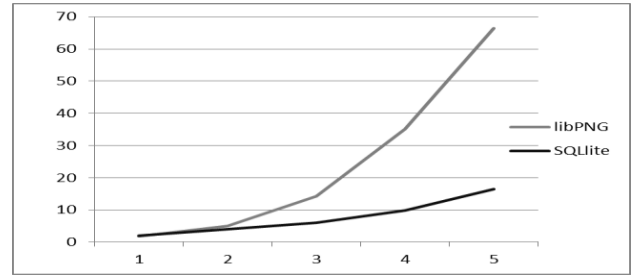

Fig. 8. Comparing libPNG and SQLite

The results described in this section have several implications for path exploration based software engineering methods. Firstly, while choosing a SMT solver for solving path feasibility queries, it is better to go with STP as it outperforms Z3. Secondly, it is better to generate the path condition formula in SMTLIB2 format as it is concise and smaller in size which performs better in case of larger formulas. And finally, the method should be optimized for both data and control flow dependencies as they together tend to generate formulas which do not scale well while solving. In the next section we illustrate this by designing a simple analysis which can help in reducing the complexity of generated path feasibility queries.

### VI. APPLICATION - CHANGE VALUE ANALYSIS (CVA)

We describe a simple program analysis which when combined with exiting compiler optimizations reduces the data and control flow dependencies. Our hypothesis is that aggressive compiler optimizations can help in efficient symbolic execution. In order to enable aggressive compiler optimizations for efficient symbolic execution we exploit undefined behaviors in programs. In particular for our analysis, we start with an initial set of variables V and a 3-point lattice (Changed, Unchanged and Undefined). Our analysis works on a program in SSA form where all data dependencies are made explicit in the code. Using the SSA form of the intermediate code we proceed as follows.

- Initially all instructions are marked "Undefined"
- All instructions which use the Value of variables from the set V are marked "Changed"
- All the instructions which depend on "Changed" instructions are marked "Changed". Other instructions are marked "Unchanged".
- We continue until a fix point is reached, which is guaranteed as in worst case all instructions will be marked changed.

We have implemented this analysis as a pass inside the LLVM compiler. We use the existing alias analysis and data flow analysis (for SSA) of LLVM. At the end of the analysis we mark all uses of "Unchanged" and "Undefined" values as a special LLVM value of "UnDef". The LLVM compiler treats "UnDef" as a non-deterministic value which enables subsequent compiler optimizations (like Dead Code Elimination) to use any value in lieu of "UnDef". This leads to a binary which is smaller and optimized for symbolic execution. The initial set of variables V is used to specify the program variables which are going to be used symbolically in the generation of path feasibility queries. The effect of this Change Value Analysis (CVA) is to systematically introduce undefined behavior in the program where we know that the behaviors is not going to be explored symbolically by the tool used to generate path feasibility queries.

We illustrate the approach with the following example. Consider a simple program with the following code written in an intermediate language of a compiler like LLVM.

IR (Intermediate Representation) Code Snippet
```
    i = 1
    j = 2
    k = 3
    n = 4
L:
    i = i + j
    l = j + 1
    j = j + 2
    if (j > n)
    return i
    else
    k = k - j
    print (l)
    goto L
```

SSA (Single Static Assignment) Form
```
    i1 = 1
    j1 = 2
    k1 = 3
    n1 = 4
L:
    i2 = i1 + j1
    l1 = j1 + 1
    j2 = j1 + 2
    if (j2 >= n1)
    return i2
    else
    k2 = k1 - j2
    print (l1)
    goto L
```

Path Feasibility Query for Return Statement is
$i1 = 1 \land j1 = 2 \land k1 = 3 \land n1 = 4 \land i2 = i1 + j1 \land l1 = j1 + 1 \land j2 = j1 + 2 \land j2 >= n1$
(8 Conjuncts)

After DCE (Dead Code Elimination)
```
    i1 = 1
    j1 = 2
    n1 = 4
L:
    i2 = i1 + j1
    l1 = j1 + 1
    j2 = j1 + 2
    if (j2 >= n1)
    return i2
    else
    print (l1)
    goto L
```

Path Feasibility Query for Return Statement is
$i1 = 1 \land j1 = 2 \land n1 = 4 \land i2 = i1 + j1 \land l1 = j1 + 1 \land j2 = j1 + 2 \land j2 >= n1$
(7 Conjuncts)

This formula contains less number of conjuncts so we see that vanilla compiler optimizations can help in improving symbolic execution. We make it better and more focused on the symbolic variables of interest (V) using the Change Value Analysis.

After CVA(Change Value Analysis) for variable 'i'
```
    i1 = 1
    j1 = 2
    k1 = 3
    n1 = 4
L:
    i2 = i1 + j1
    l1 = * + 1
    j2 = * + 2
    if (j2 >= *)
    return i2
    else
    k2 = * - j2
    print (l1)
    goto L
```

Here * represents a non-deterministic value ("UnDef" in LLVM). During subsequent phases of compilation * may take any value suitable for a given optimization.

DCE After CVA
```
    i1 = 1
    j1 = 2
L:
    i2 = i1 + j1
    return i2
```

Path Feasibility Query for Return Statement is
$i1 = 1 \land j1 = 2 \land i2 = i1 + j1$
(3 Conjuncts)

This example shows that CVA can help reduce the complexity of the generated path feasibility queries. In experiments we have used CVA followed by aggressive existing compiler optimization in LLVM (in particular Constant Propagation, Global Value Numbering, Sparse Constant Conditional Propagation, Loop Deletion, Tail Call Elimination, Loop Invariant Code Motion, Dead Instruction Elimination, Dead Store Elimination and Dead Code Elimination). We present the results of our analysis on the SIR Programs in Table VI. We used the input variables of each program as the initial set of variables V for the analysis.

TABLE VI. CHANGE VALUE ANALYSIS FOR SIR PROGRAMS

| Program | LoC | Constraints | Constraints (CVA) | Time | Time (CVA) |
|---|---|---|---|---|---|
| tcas | 173 | 848 | 601 | 43.7 | 24.2 |
| schedule2 | 374 | 960 | 821 | 78.4 | 34.6 |
| replace | 564 | 264 | 219 | 53.9 | 39.7 |
| totinfo | 565 | 256 | 210 | 24.7 | 11.8 |
| print_tokens2 | 570 | 632 | 632 | 180.9 | 78.5 |
| space | 6199 | 100 | 91 | 82.6 | 52.5 |
| grep | 10068 | 512 | 56 | 55.3 | 19.3 |
| flex | 10459 | 576 | 340 | 180.5 | 101 |
| sed | 14427 | 144 | 17 | 13.9 | 7.5 |

In each of the experiments with SIR programs we used Fuzzgrind with a binary compiled using LLVM without CVA and compare it with another one compiled after enabling CVA. From our evaluation we find that the time taken for solving the generated path feasibility queries in reduced by 48% (on an average). The generated formulas are also much simpler as we see a reduction in the number of constraints (30% on average).

## VII. THREATS TO VALIDITY

We discuss the internal and external threats to the validity of this study. Our experiments are carried with three kinds of programs, GDSL, SQLite and libPNG. But the path conditions are generated from the assembly code and not the source code of the program. The results obtained may be influenced by this, as we did not find any difference in performance between different sections of the same path in the program. When looking at the program at assembly code it looks a lot more uniform due to limited instructions available and simple conditional constructs. This can be mitigated by using a tool which generates path conditions at source code level, but as we saw in section II it is not always possible to do so due to external calls.

Also in general there can be an unbounded number of paths in a program we cannot generate path feasibility queries for all of them. We use different inputs to direct the execution of a program towards different paths. This represents a subset of the total paths in the program and those paths may give different results when used for path feasibility queries. We attempt to cover the normal execution of the program in this study; other paths like exceptional flow may lead to different results. The inter-program performance analysis is done using only two different programs, other programs may lead to different results. However we chose the two programs with very different uses and expected behavior, a database and an image manipulation library. Since, these programs represent some of the popular real world C programs which are widely used and have been in development for quite some time they are expected to have typical C programming constructs and idiosyncrasies; hence the results obtained are likely to be useful.

Now we discuss some external threats which affect the generalizations we reach and the recommendations we give based on those results. Some particular kinds of programs like numerical libraries may have sections in a particular path in the program (corresponding to intensive computation) which give problems for checking path feasibility queries. Even though this is certainly possible, our path feasibility queries are based on assembly code which is more likely to give results which can be generalized to all programs. The ratio we use to characterize a given program may not be useful at all for programs with different unrelated paths. This is certainly a weak point of this study, however even for those kinds of programs if the paths are not shared at all between different executions we can treat such a group of paths as representing a totally separate behavior and analyze it as another sub program.

Another possible obstacle in generalizing the result can be that the path feasibility queries generated by Bitblaze are very inefficient and another tool may give better results. While this is a certain possibility and one can argue that the method used in KLEE where full path condition formula is not generated by a weakest precondition calculation but by symbolic execution is a better way to check for path feasibility queries. The KLEE-like method is useful for test case generation but not for some other software engineering methods like program debugging and trace comparison, where we need the full path condition. Our study shows the difficulties for such kind of methods which use path feasibility queries.

There is a tradeoff to be made while trying to address threats to external and internal validity, consider the choice of using assembly code for path condition generation. This allows us to generalize the results but we may miss some issues related to different constructs used in the source code. Our choice is (based on Bitblaze) to use assembly code but generate full path conditions to create a balance between generalization of results and discovering various constructs that give problems for SMT Solvers.

## VIII. RELATED WORK

We are not aware of any existing work on empirical studies of path feasibility queries as generated by symbolic execution engines. The existing SMT competition [9] benchmarks are based on formulas generated using verification conditions. Recent work [13] comparing verification condition generation and symbolic execution has shown that efficient symbolic execution can lead to faster verification. Phang et al. describe a system [15] which mixes type checking with symbolic execution which is more efficient than exclusive symbolic execution. Our work can help designers of such new analysis based on symbolic execution in choosing the right solvers, format of queries and compiler optimization in order to make their analysis more efficient.

Existing work [14, 16] on Multi-path analysis aims to reduce some of the time taken by exploration during symbolic execution. In contrast we have focused on reducing the complexity of each generated path feasibility query. The Change Value Analysis (CVA) proposed in this paper can enhance existing multi-path analysis by statically marking irrelevant parts of program as undefined behavior. We utilize existing compiler optimizations to generate a binary which is more suitable for symbolic execution. There are other specialized program analysis based approaches [17, 18] for symbolic execution. These methods use abstractions and/or concretizations for relevant portions of program which are deemed hard for symbolic execution. In our work we have avoided it by making use of undefined behaviors in LLVM compiler. Also the technique (CVA) presented in this paper is much simpler and uses existing optimizations.

The methods which aim to improve on constraint solving [20] for symbolic execution are orthogonal to our work. As we show for the SIR benchmark CVA can reduce the number of constraints in a generated formula which will also benefit other optimizations which are implemented in constraint solvers. Specialized procedures like [19] can model the advanced programming language structures like Strings directly. Our work is limited to the theory of bit vector and arrays as it forms the core of most existing symbolic execution engines (like Klee, Bitblaze and Fuzzgrind).

## IX. CONCLUSIONS AND FUTURE WORK

We carried out an empirically study of path feasibility queries generated from 3 real world C programs – GDSL, SQLite and libPNG. We found that for path feasibility queries the SMT solver STP performs better than Z3 by an order of magnitude. We showed that SMTLIB2 format is concise and leads to shorter formulas. We discovered a relation between control and data dependencies along a path and the scalability of the path feasibility query generated along that path. Our results suggest that path feasibility queries generated by Bitblaze contain redundancy and can be improved by outputting the formula in SMTLIB2 format directly.

While designing path exploration based software engineering methods we need to consider both data and control dependencies in order to scale the method for larger programs. The formulas [12] we generated for path feasibility queries can be used a benchmark to supplement existing benchmarks like ones used in [9]. This can be used for comparing optimizations performed in various SMT Solvers. As an application of our study we designed a new program analysis (Change Value Analysis). We showed that our analysis reduces the time taken to solve path feasibility queries generated from SIR benchmark of programs by 48%. It also leads to simpler and smaller formulas (30% on average) to be generated.

For future work we would like to develop the results of section V C into a complete empirically model, which can be used to predict the performance of SMT solvers for different programs. That would enable us to apply the results of this study to other programs directly. We would also like to see if it is possible to use other theories like uninterpreted functions to capture a path condition at source code level which doesn't suffer from the problems of external calls and system commands.